\documentclass[aps,twocolumn,amssymb,amsfonts,amsmath,showpacs,final,a4paper,superscriptaddress]{revtex4-2}

\usepackage{float}

\usepackage[dvips,final]{graphicx}
\usepackage{dcolumn}
\usepackage{bm}
\usepackage[latin1]{inputenc}
\usepackage{textcomp}
\usepackage{amsmath}
\usepackage{color}

\begin{document}
\title{Femtosecond electron transfer dynamics across the D$_2$O/Cs$^+$/Cu(111) interface: The impact of hydrogen bonding}

\author{John Thomas}
\altaffiliation{Current address: Max-Born-Institute, 12489 Berlin, Germany}
\affiliation{Faculty of Physics and Center for Nanointegration (CENIDE), University of Duisburg-Essen, Lotharstr.~1, 47057 Duisburg, Germany}

\author{Jayita Patwari}
\affiliation{Faculty of Physics and Center for Nanointegration (CENIDE), University of Duisburg-Essen, Lotharstr.~1, 47057 Duisburg, Germany}
\affiliation{Physical Chemistry I, Ruhr-University Bochum, Universit\"{a}tsstr.\,150, 44801~Bochum, Germany}

\author{Inga Langguth}
\affiliation{Physical Chemistry I, Ruhr-University Bochum, Universit\"{a}tsstr.\,150, 44801~Bochum, Germany}

\author{Christopher Penschke}
\affiliation{Department of Chemistry, University of Potsdam, Karl-Liebknecht-Str.\ 24-25, D-14476 Potsdam-Golm, Germany}

\author{Ping Zhou}
\affiliation{Faculty of Physics and Center for Nanointegration (CENIDE), University of Duisburg-Essen, Lotharstr.~1, 47057 Duisburg, Germany}

\author {Karina Morgenstern}
\affiliation{Physical Chemistry I, Ruhr-University Bochum, Universit\"{a}tsstr.\,150, 44801~Bochum, Germany}

\author{Uwe Bovensiepen}\email[] {uwe.bovensiepen@uni-due.de}
\affiliation{Faculty of Physics and Center for Nanointegration (CENIDE), University of Duisburg-Essen, Lotharstr.~1, 47057 Duisburg, Germany}
\affiliation{Institute for Solid State Physics, The University of Tokyo, Kashiwa, Chiba 277-8581, Japan}


\begin{abstract}
Hydrogen bonding is essential in electron transfer processes at water-electrode interfaces. We study the impact of the H-bonding of water as a solvent molecule on real-time electron transfer dynamics across a Cs$^+$-Cu(111) ion-metal interface using femtosecond time-resolved two-photon photoelectron spectroscopy. We distinguish in the formed water-alkali aggregates two regimes below and above two water molecules per ion. Upon crossing the boundary of these regimes, the lifetime of the excess electron localized transiently at the Cs$^+$ ion increases from 40 to 60~femtoseconds, which indicates a reduced alkali-metal interaction. Furthermore, the energy transferred to a dynamic structural rearrangement due to hydration is reduced from 0.3 to 0.2~eV concomitantly. These effects are a consequence of H-bonding and the beginning formation of a nanoscale water network. This finding is supported by real-space imaging of the solvatomers and vibrational frequency shifts of the OH stretch and bending modes calculated for these specific interfaces.
\end{abstract}

\maketitle
\section{Introduction}

Ion-solvent interaction at the nanoscale and electron transfer across heterogeneous interfaces are the key factors for energy storage and energy conversion applications \cite{Regan_1991, Fujishima_1972}. Since the last decade, a number of experimental and theoretical \cite{Roy_2018, Lee_2017, Bragg_2010}  investigations have been performed to understand the structural and dynamic properties of such interfaces. Among the widely investigated solvents, water has received special attention because of its wide abundance and relevance of hydration dynamics in various fields like ion transport through membrane channels and nano-confined environments \cite{swanson_2007,kratochvil_2016,cao_2010} to ion diffusion in energy conversion systems \cite{pean_2015,ong_2015}. In living cells, water is considered to be a versatile and adaptive component because it combines the properties of a small molecule solvent with hydrogen bonding \cite{ball_2007}. Alkali-water interactions at metal interfaces involve a complex interplay of several mechanisms such as hydrogen bonding, van der Waals interaction, and image charge effects \cite{ paz_2021}. Among the interactions relevant for the polar protic solvent water, H-bonding is fundamental since it influences the electronic and geometric structure and, consequently, the hydration dynamics of the complex multi-component system \cite{Nibbering_2007}.

Hydrogen bonding is key in the hydration process and ubiquitous in nature \cite{xantheas_dunning_1993,Nibbering_2007}. Being stronger than most other intermolecular interactions, hydrogen bonds can be decisive for the structural and dynamical properties of nanoscale systems. It allows, for example, to facilitate tuning biological electron transfer rates \cite{Rege_1995}. Recently, this concept has been extended to photocatalytic reactions \cite{Breg_2021} and electrochemical interfaces \cite{Chen_2022} where H-bonding has been tuned to regulate the electron transfer efficiency to benefit the particular application. Thus, at interfaces the influence of H-bonding on electron transfer is  scientifically and technologically important.

Vibrational spectroscopy along with different theoretical models provided significant insight into the influence of hydrogen bonding in solvent networks. Upon H-bonding the O-H stretch vibration frequency red-shifts by about 10~\% and the bending modes exhibit a weaker blue-shift \cite{Nibbering_2007}. H-bonding influences each of the possible motion of water molecules in a solvent network such as stretching, in-plane bending, and libration \cite{Cowan_2005}. However, real-time information about the dynamics is challenging to obtain experimentally \cite{Yang_2021}. Time-resolved spectroscopy facilitated probing different dynamical processes on femto- to picoseconds which are relevant in H-bonding \cite{Nome_2010}. Since such studies were performed in the bulk, the contribution arising from individual water molecules could not be derived due to the large number of water molecules. At femtosecond timescales, water molecules exhibit microscopic dynamics which include hydrogen bond strengthening, librational motions, and mixed stretch-bend motions \cite{Yang_2021, Cowan_2005, Ramasesha_2013}. At interfaces, hydrogen-bonded nanostructures are highly dynamic. For instance, one molecule of a water dimer rotates around a second one, as calculated on Pd(111) \cite{Ranea_2004} and experimentally revealed on Pt(111) \cite{Motobayashi_2008}. The water dimer was imaged here as a six-fold flower-like protrusion, reflecting the six symmetry-equivalent positions of the rotating molecule around the static one. This interpretation was put forward for water dimers on Cu(111) \cite{Bertram_2019} and water solvating a Na ion on Cu(100) \cite{Shiotari_2018}. Both are imaged as round protrusions despite their elongated geometry.

We analyzed the interactions governing the structure of the D$_2$O/Cs$^+$/Cu(111) interface  \cite{penschke_2023} and concluded that H-bonding among D$_2$O is dominant. Energy stabilization is characterized through a split-off electronic state, contrary to the case of Xe as a non-polar solvent \cite{thomas_2021}. Although we have been investigating the variation of hydration dynamics for different alkali ions \cite{meyer_2015}, a dynamic effect of H-bonding on the electron transfer dynamics remained elusive.

Here, we investigate the electron transfer dynamics of alkali ion Cs$^+$ adsorbed on a Cu(111) electrode in the presence of water as a co-adsorbate using femtosecond time-resolved two-photon photoelectron spectroscopy (2PPE) combined with low-temperature scanning tunneling microscopy (STM). We analyze the variation of excited state lifetime and energy stabilization due to energy transfer to the solvent upon variation of the number of water molecules per ion $\rho$. We identify in both observables different regimes without and with H-bonding being present. A sharp transition between the regimes is assigned the change from singly- to triply-solvated Cs$^+$ ions.

\begin{figure}[t]
    \centering
    \includegraphics[width=0.99\columnwidth]{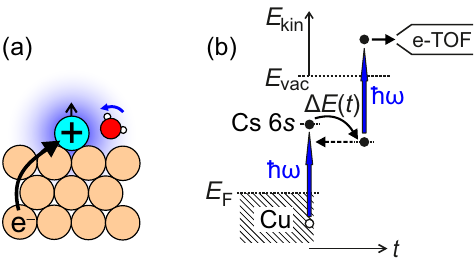}
    \caption{(a) Illustration of this photo-excited electron transfer in real space from the Cu bulk to the adsorbed ion and potential response of the surrounding water to the modified ion charge. (b) Scheme of photo-excited electron transfer from the Cu surface to the Cs $6s$ orbital of the alkali ion adsorbed on the Cu(111) surface and time-resolved two-photon photoelectron spectroscopy. The kinetic energy of the photoelectron $E_{\mathrm{kin}}$ is analyzed by an electron time-of-flight spectrometer (e-TOF).}
    \label{fig:fig1}
\end{figure}

\section {Experimental Details}
For the experimental set-up, the Cu(111) single-crystal serves as a surface and an electrode for electron transfer. It is kept in ultrahigh vacuum (UHV) at a base pressure $2 \cdot 10^{-10}$~mbar ($5 \cdot 10^{-10}$~mbar) for 2PPE (STM) experiments. It is prepared by ion sputtering cycles using Ar at $2 \cdot 10^{-5}$~mbar ($3 \cdot 10^{-5}$~mbar) pressure at 1.5~keV 1.3~keV ion energy for 10--30~min and subsequent annealing at a temperature $T = 600$~K {(893 $\pm$ 1~K)} for 10--20~min. The investigated Cs$^+$ ions are chemisorbed on Cu(111) at $T = 200$~K {(249 $\pm$ 4~K)}. They are generated as neutral atoms from commercial getter sources (SAES getters). The Cs coverage in monolayers (ML) $\Theta_{\mathrm{Cs}}$ is defined with respect to a closed-packed $(2\times 2)$ layer on Cu(111). $\Theta _{\mathrm{Cs}}$ is determined {in the 2PPE experiments} by measuring the change in work function $\Phi=E_{\mathrm{vac}}-E_{\mathrm{F}}$ of the surface upon Cs adsorption as described in Ref.~\cite{lu_1996}; $E_{\mathrm{vac}}$ and $E_{\mathrm{F}}$ are the vacuum energy and the Fermi energy of Cu(111), respectively. Liquid water (D$_2$O) is stored in a test tube as a part of the gas system. It is connected to a main and a dosing reservoir. After purification of the water in the test tube through several freeze-pump-thaw cycles water is introduced at a pressure of 0.1~mbar into the main reservoir. Subsequently, it is expanded into the dosing reservoir and through a pinhole of 50~$\mu$m diameter into the UHV for a controlled dosing time at constant pressure, which are both proportional to the desired water coverage on the Cu(111) surface. The Cu(111) surface is kept in UHV at a temperature of 80~K in front of a stainless steel tube which guides the water vapor in UHV from the pinhole to the sample. The D$_2$O coverage $\Theta_{\mathrm{D_2O}}$ is quantified by temperature programmed desorption (TPD) and measurement of the change in $\Phi$ \cite{bovensiepen_2003}. The amount of dosed water was calibrated previously by water adsorption on Ru(001) in terms of bilayers \cite{bovensiepen_2003}. One BL refers to a hexagonal layer with 2 molecules per 3 surface atoms, in which neighboring molecules are displaced by 96 pm vertically. Thus, 1~BL refers to a closed single molecular layer of water ice. The determination of $\Phi$ by two-photon photoemission spectroscopy (2PPE) allows the determination of mass equivalent coverages as fractions of 1~BL \cite{bovensiepen_2003}. With this analysis, we obtain the number of respective ions and molecules per surface area, which gives the spatially averaged ratio $\rho$ of D$_2$O molecules to Cs$^+$ ions on the Cu(111) surfaces. We varied the coverage ratio  $\rho$ from 1 to 7 in our experiments which corresponds Cs-water clusters of a few nanometer sizes on Cu(111).

To analyze the interaction of D$_2$O with Cs$^+$/Cu(111), we monitor the dynamic response of the interface upon resonant, photo-induced electron transfer to the Cs $6s$ derived wave function with an increasing number of water molecules coadsorbed onto the surface in femtosecond time-resolved 2PPE, as illustrated in Fig.~\ref{fig:fig1}.

In the 2PPE experiment, two photons of energy $\hbar \omega=3.1$~eV, which is below the work function $\Phi$, are absorbed and emit one photoelectron with kinetic energy $E_{\mathrm{kin}}$. As illustrated in Fig.~\ref{fig:fig1}b, one photon from a first femtosecond laser pulse excites resonant electron transfer from Cu(111) to the unoccupied Cs $6s$ state of the adsorbed Cs$^+$ ion \cite{bauer_1997, petek_2000}. A second photon from a second laser pulse at a time delay $t$ generates the photoelectrons analyzed in an electron time-of-flight spectrometer. Femtosecond laser pulses at 1.55 eV photon energy and 40~fs pulse duration are generated in a commercial Ti:sapphire amplifier (Coherent RegA 9040) and frequency doubled in BaB$_2$O$_4$ to $\hbar \omega=3.10$~eV. For details of the experimental setup see \cite{sandhofer_2014}.

Density functional theory (DFT) calculations using PBE \cite{perdew_1996} with D3 dispersion corrections \cite{grimme_2010,grimme_2011} were performed using VASP \cite{kresse_1996a,kresse_1996b}. The Cu(111) surface is modelled by a three-layer 5$\times$5 unit cell. Additional details on the DFT calculations, as well as the cluster structures, are presented in the Supplemental Material of Ref.~\cite{penschke_2023}. For the present work, numerical vibrational frequency calculations using central differences with a displacement of 1.5 pm were performed.

For the STM experiment, the water is purified in a glass tube by several freeze-pump-thaw cycles before introducing it into a dedicated molecule deposition chamber via a leak valve. At a vapour pressure of approx.\ $5\cdot 10^{-5}$~mbar it is expanded into the preparation and STM chamber, reducing the pressure by several orders of magnitude. The water is deposited onto the sample situated within the cold shields surrounding the STM for approx.\ 1.5~s. During deposition, the sample temperature rises slightly to less than 13~K. The D$_2$O coverage 
of $0.34$~\% bilayer (BL) is determined from pure water structures on the pristine metal resulting from the same procedure.

\begin{figure*}[t]
    \centering
    \includegraphics[width=0.8\textwidth]{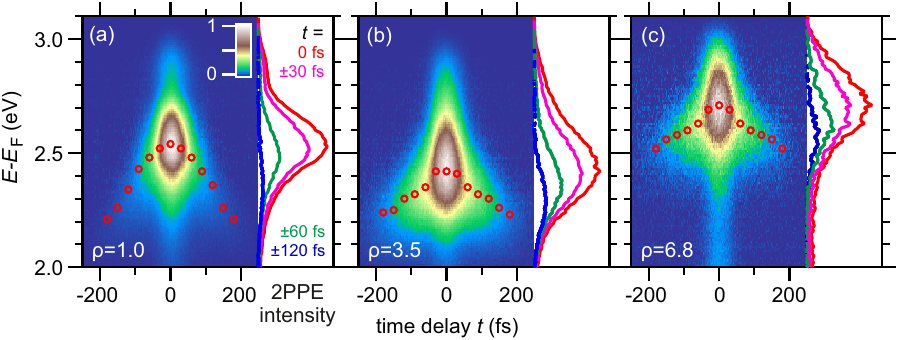}
    \caption{The panels (a-c) depict measured data sets of time-resolved 2PPE intensity autocorrelations with increasing coverage ratio $\rho$ (number of water molecules per adsorbed ion) after subtraction of a time-independent background signal. In each individual panel, the intensity is shown in a false color representation as a function of time delay and intermediate state energy $E-E_{\mathrm{F}}$ at left and as spectra at selected time delays after averaging negative and positive time delays at right. The open red circles indicate the energy of the intensity maxima at the respective time delay. The data were recorded at a temperature of 80~K.}
    \label{fig:fig2}
\end{figure*}

\section {Results and Discussion}

Fig.~\ref{fig:fig2} shows time-resolved 2PPE intensity autocorrelation measurements for $\rho=1.0, 3.5,$~and~6.8. The pronounced peak at $t=0$ represents the photo-induced electron transfer resonance from electronic states in Cu near $E_{\mathrm{F}}$ to the $6s$ derived state of Cs$^+$/Cu(111) \cite{gauyacq_2007}. Two effects are recognized immediately. First, the energy of the electron transfer resonance shifts non-monotonously with $\rho$. This behavior results from the variation of the static electronic and geometric structure with $\rho$, see \cite{penschke_2023}. Second, the time-dependent dynamics depend on $\rho$. For smaller $\rho$ the intensity decays faster and the overall peak energy shift is higher compared to larger $\rho$, see the red circles in ~Fig.~\ref{fig:fig2}. These effects represent water-induced variations in the interaction of the Cs $6s$ derived wave function with the electronic states in Cu(111) and energy transfer in response to the nuclear motion, respectively.

\begin{figure}[h!]
    \centering
    \includegraphics[width=0.780\columnwidth]{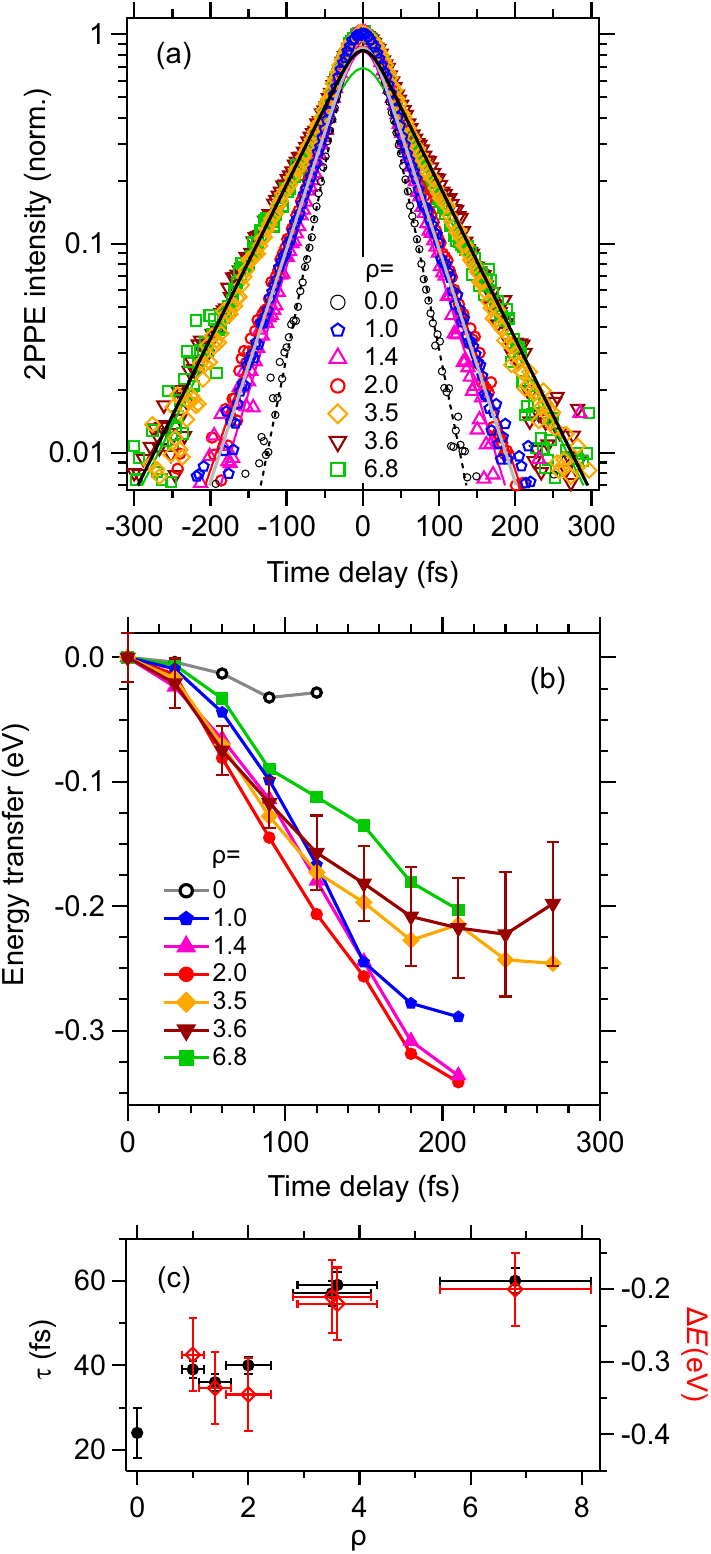}
    \caption{(a) The energy integrated time-dependent intensity of the 2PPE autocorrelation for different coverage ratio $\rho$ is indicated by different symbols. $\rho=0$ is a reference measurement for bare Cs$^+$/Cu(111) without coadsorbed water. Lines are the results of modeling the single exponential decay with relaxation time $\tau$ convoluted with a Gaussian to describe the autocorrelation of the UV laser pulses. The gray line describes the data for $\rho=1.0$ and the black line $\rho=3.5$. The color of the further lines indicates $\rho$ according to the legend. (b) time-dependent energy transfer from the electronic to the nuclear degree of freedom determined from the energy shift of Cs $6s$ resonance in the time-resolved 2PPE spectra for different coverage ratios $\rho$. Data for negative and positive $t$ are averaged. The vertical error bars indicated for $\rho=3.6$ are representative of all $\rho$. (c) Filled black circles represented the relaxation times (left axis) as a function of $\rho$ determined by the single exponential decay fit to the data in (a). Open symbols in red depict the energy transfer (right axis) of the Cs 6s resonance to nuclear motion taken from the data shown in (b) at a time delay of 210~fs.}
    \label{fig:fig 3}
\end{figure}

For further analysis of the transient electron population, the time-dependent 2PPE intensity was integrated within an energy interval of 1~eV to include the entire electron transfer peak in the 2PPE spectrum. The intensity is normalized at $t=0$ for different $\rho$. The results are depicted in Fig.~\ref{fig:fig 3}a. The obtained time-dependent population traces exhibit two regimes ($\rho>2$ and $\rho<2$) with identical behavior within each regime. While in both the regimes, the decay is clearly slower than for bare Cs$^+$/Cu(111) at $\rho=0$, the decay time increases with $\rho$. We fit these experimental data with a single exponential relaxation convoluted with a Gaussian which represents the laser pulse autocorrelation, see the solid lines in Fig.~\ref{fig:fig 3}a. This fit describes the single exponential relaxation observed for $|t|>30$~fs very well. For $|t|<30$~fs the fit deviates from the experimental data, which is explained by coherent 2PPE processes that do not lead to a population of the electron transfer resonance and the Cs $6s$ state. A population build-up requires phase-breaking events which lead to an incoherent 2PPE contribution \cite{petek_ogawa_1997,weinelt_2002}, which follows the exponential decay with a decay time $\tau$.

Fig.~\ref{fig:fig 3}c represents the obtained values for $\tau$ as a function of $\rho$. The fit results for $\tau$ upon adsorption of D$_2$O indicate an increase in $\tau$ to $\sim40$~fs for $\rho<2$ and to $\sim60$~fs for $\rho>2$ in two distinctive steps. As discussed by Gauyacq et al. \cite{gauyacq_2007} the decay time measured in time-resolved 2PPE intensity is due to elastic wavepacket propagation from the localized Cs $6s$ wavefunction to delocalized Bloch states in Cu(111). An increase in lifetime upon water coadsorption to Cs$^+$/Cu(111) as observed in Fig.~\ref{fig:fig 3}a implies that already few water molecules weaken the electronic Cs-Cu interaction considerably by screening. While in dielectric environments, like Xe \cite{thomas_2021}, screening can be strong due to the large electron density ($Z=54$), in water with 18 electrons, the molecular dipole will lead to considerable screening by its reorientation in the hydration response. Note that  on the same Cs$^+$/Cu(111) surface, we observed a six-time increase in $\tau$ for Xe co-adsorption compared to a 2.5 times increase in the case of D$_2$O. We attribute the increase of $\tau$ in the case of water coadsorption to dipolar screening rather than to dielectric screening. The occurrence of two distinct structural regimes above and below $\rho=2$ indicates that besides the dipolar character, the static structure plays an important role. It is plausible that the strength of dynamic, dipolar screening effects indeed depends on the orientation of the molecular dipole and its response to a change of the charge count of the alkali ion as in these experiments induced by resonant charge transfer on the femtosecond timescale.

With the time-dependent energy shift $\Delta E$ of the electron transfer state we analyze the hydration dynamics. Its sign is assumed to be negative upon energy transfer from the electronic system to the solvent. We determine the time-dependent shift of the peak maxima of the 2PPE spectra, see Fig.~\ref{fig:fig2}, for different $\rho$. The results are reported in Fig.~\ref{fig:fig 3}b. The two structural regimes below and above $\rho=2$ which were mentioned above are also affecting this dynamic observable. For $0<\rho \leq 2$ the energy $\Delta E = 0.3$~eV is transferred to hydration modes within 210~fs. For $2<\rho<7$ the energy transfer is 0.2~eV within the same time window. Fig.~\ref{fig:fig 3}c further depicts the amount of total energy stabilization of peak energy as a function of $\rho$. It also depicts two different regimes below and above $\rho=2$ which resembles the lifetime dependence on $\rho$. The longer lifetimes $\rho>2$ allow to analyze the energy gain up to larger time delays than for $\rho<2$.

\begin{figure}[h]
\centering
\includegraphics[width=0.9\columnwidth]{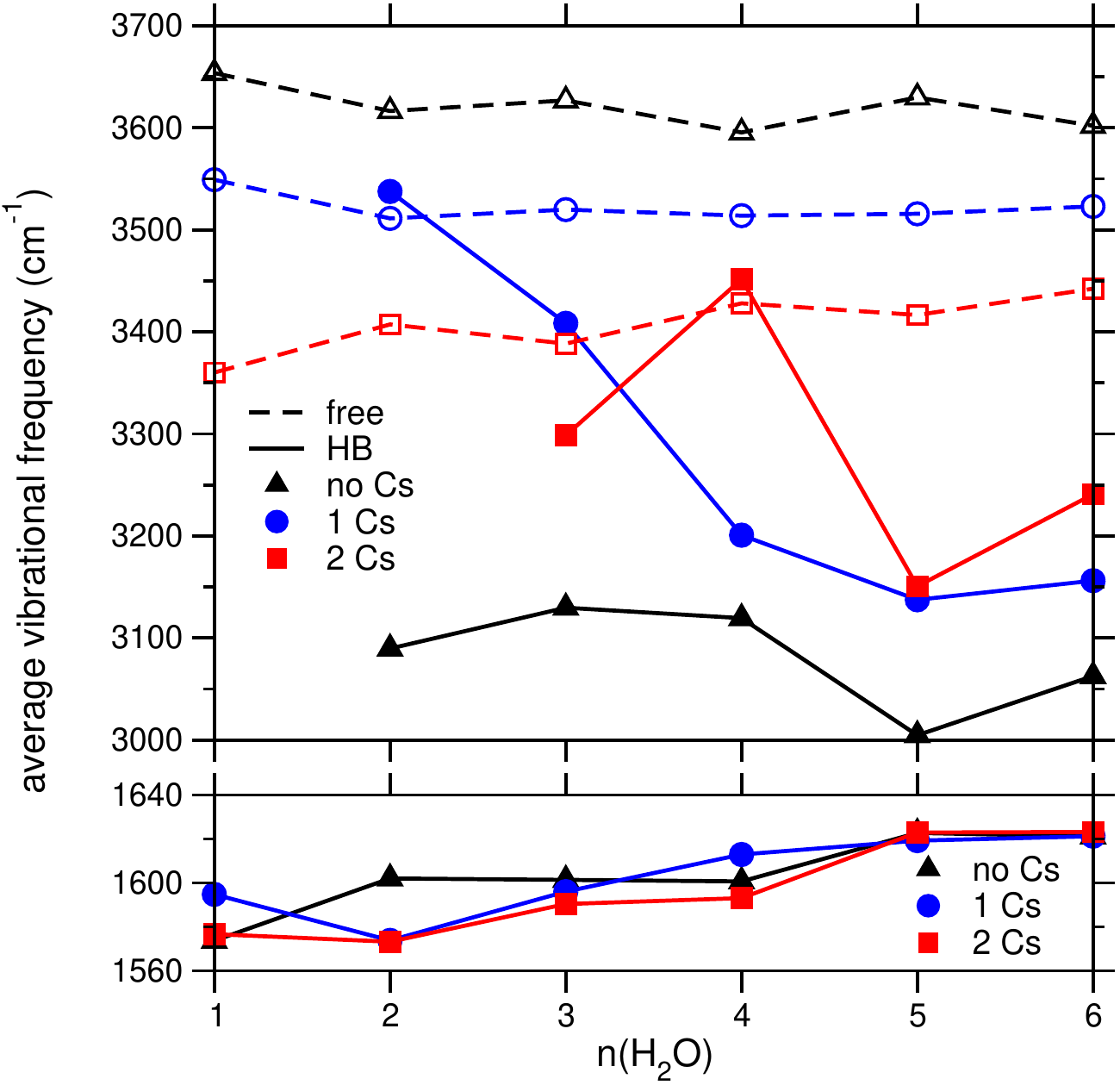}
    \caption{Harmonic vibrational frequencies of the water molecules in Cs$^+$-H$_2$O clusters on Cu(111), averaged over all respective modes, as a function of water coverage. The upper panel shows the O-H stretch vibrations, divided into hydrogen-bonded (HB, filled symbols) and free (open symbols) OH groups. The lower panel depicts bending modes.}
    \label{fig:fig-t}
\end{figure}

We explain the two distinct regimes in the electron lifetime and the energy stabilization by a different solvent response upon electron transfer in the absence and presence of H-bonding below and above $\rho=2$, respectively. In our earlier work, we analyzed the geometric and the electronic structure \cite{penschke_2023}. We concluded a peculiar inside-out solvation structure for the higher water coverage and appearance of a second electronic state lower in energy at $\rho=2$ which are explained by competing interactions at the interface dominated by H-bonding.

In order to validate the onset of H-bonding on $\rho=2$ further, we calculated the harmonic vibrational frequencies of the water molecules in the solvated species  Cs$^+$-H$_2$O on Cu(111) investigated in experiments using DFT. Focusing on the modes above 1000~cm$^{-1}$, there are two main types of vibrations: O-H stretch vibrations (above 3000~cm$^{-1}$) and H-O-H bending vibrations (around 1600~cm$^{-1}$). The O-H bonds and the corresponding stretching vibrations can be further divided into hydrogen-bonded (HB) and free, or non-HB, OH groups. The non-HB frequencies are higher than the HB ones (see Fig.~\ref{fig:fig-t}), and they are mostly independent of the water coverage. In contrast, HB frequencies decrease with increasing water coverage by more than 10\%. While the free OH modes are systematically red-shifted with increasing Cs coverage, the HB stretch frequencies are mostly blue-shifted. Bending modes are slightly blue-shifted with increasing water coverage, but they are independent of the Cs coverage. These frequency shifts due to increasing water coverage agree very well with previous experimental and theoretical work upon the onset of hydrogen bonding \cite{Nibbering_2007,dunn_2006}. Note that these trends are not strongly affected by the presence of Cs, even if the stretch frequencies themselves depend on the Cs coverage. We do not aim to accurately reproduce experimental vibrational frequencies with these calculations. Instead, they are meant to show how the water and Cs coverage qualitatively affect water vibrations. There are several factors affecting the accuracy of the calculated frequencies, such as the use of an approximate density functional, the neglect of anharmonic effects, and the use of H$_2$O instead of D$_2$O, which are discussed in the literature \cite{dunn_2006,kesharwani_2015,ceriotti_2016}.

A naive expectation of the energy transfer to the solvent could be that the more water molecules participate in the solvation process more energy can be transferred. However, our results show that the contrary is correct. As depicted in Fig.~\ref{fig:fig 3}b the energy transfer measured for $\rho \le 2$ is 50\% larger than for more than 2 water molecules per ion. Considering that at $\rho=2$ hydrogen bonding sets in --  which is not surprising since at least two water molecules are required to form a mutual hydrogen bond --  the reduced energy transfer to the solvent for $\rho>2$ is explained by a competing interaction. Hydrogen bonding leads to formation of a water network which results in a reduced mobility compared to individual decoupled water molecules and explains our observation. The competition between energy transfer to the solvent and hydrogen bonding becomes very clear if the energy transfer $\Delta E$, which is decreasing as a function of $\rho$ from 0.3 to 0.2~eV, is compared to typical values of hydrogen bonding of 0.21~eV for the water dimer \cite{xantheas_dunning_1993}. Finally, both energies are very similar and may compete. The effect of water on the electronic lifetime as a function of $\rho$ for hydrated Cs$^+$ ions on Cu(111) is the opposite in this context. For a higher number of water molecules within the nanocluster, the water network is less mobile and might screen the ion-metal interaction better compared to a smaller number of water molecules which is highly mobile. More water hence supports the localization of the electron to the ion and increases the lifetime.

\begin{figure}[t]
    \centering
    \includegraphics[width=0.99\columnwidth]{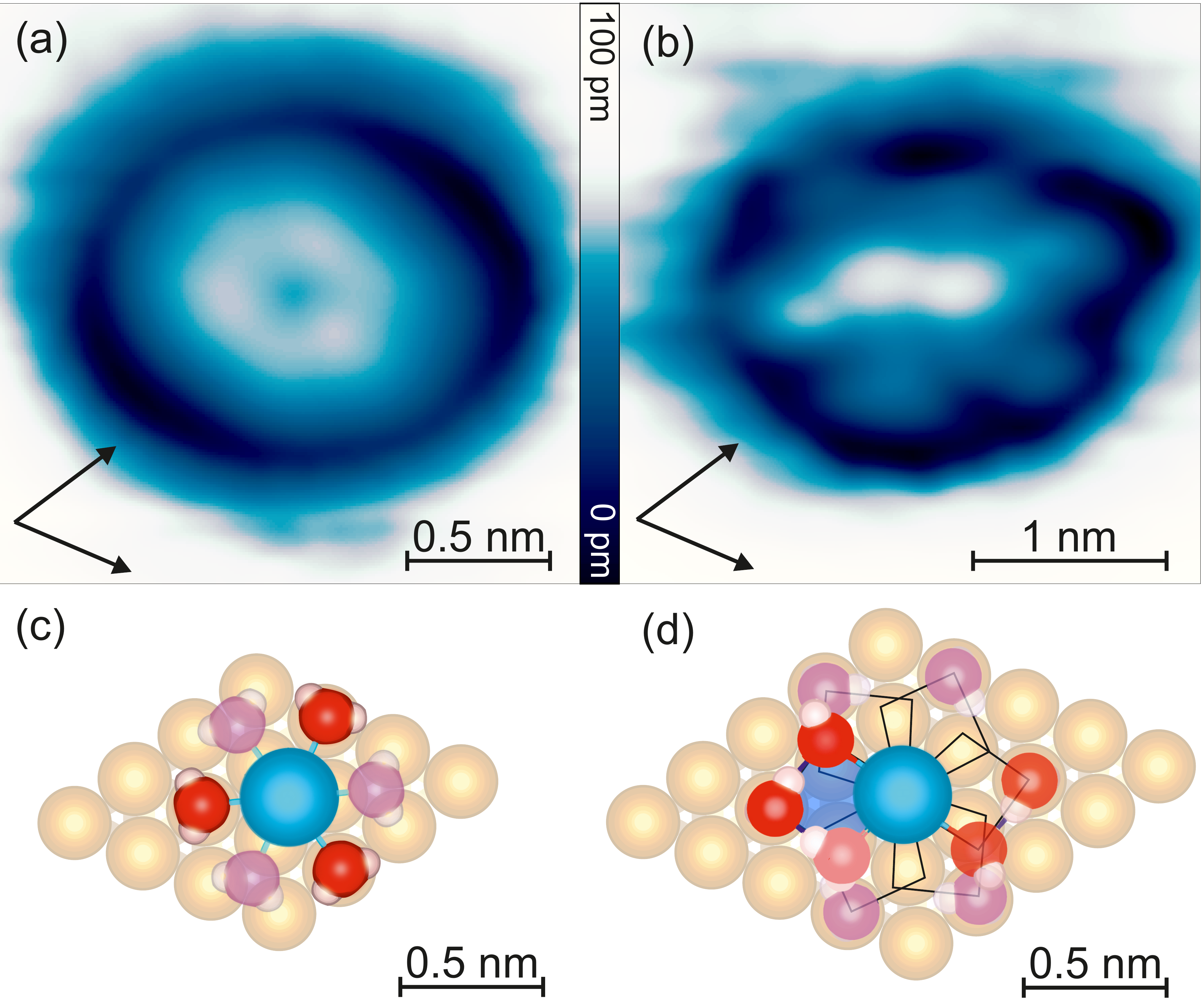}
    \caption{D$_2$O-Cs$^+$ solvatomers on Cu(111): (a,b) STM images of singly-solvated Cs$^+$ of three-fold symmetry (a) and triply-solvated Cs$^+$ of six-fold symmetry (b) on different scales as marked; arrows mark two of the $\langle 110 \rangle$ surface directions (c,d) Scheme of six rotamers of singly-solvated Cs$^+$ (c) and triply-solvated Cs$^+$ (d) around the Cs$^+$ at hollow (c) and on-top site (d). Optimized adsorption structure from \cite{penschke_2023}, SI; red and magenta oxygen atoms mark D$_2$O molecules yielding larger and smaller apparent heights in the STM images, respectively. For clarity, the three water molecules are displayed only for one of the rotamers in (d), two in solid red and one in half-transparent red framing a blue-shaded rhombus, and the others are sketched by rhombi with only the outer water molecule displayed. The STM images were recorded at (a) I~=~64~pA, V~=~12~mV and (b) I = 150~pA,  V = 9~mV.} 
    \label{fig:fig 6}
\end{figure}

To support our claim and elucidate the sharp transition, we performed low-temperature STM measurements of Cs$^+$-D$_2$O clusters formed on Cu(111) at 13~K under vast Cs excess. Imaging at very small voltages yields two species that differ in size and symmetry.
The main species (89~\%) is torus-like with a fine structure of three-fold symmetry (Fig.~\ref{fig:fig 6}a). The distance between the maxima  on the torus is 0.44~nm. The second most common species (11~\%) has flower-like six-fold symmetry (Fig.~\ref{fig:fig 6}b). The maxima of both species are aligned along the $\langle 112 \rangle$ surface directions. 

Lateral size and a large excess of Cs suggest that the smaller species is a singly solvated Cs$^+$, imaged at six centrosymmetric positions around an anchoring center, the Cs$^+$. Such a time-averaged STM image of six rotamers resembles that of water dimers on Pt(111) \cite{ Motobayashi_2008}, explaining also the round shape of water dimers on Pd(111) \cite{Ranea_2004} and Cu(111) \cite{Bertram_2019}. The three-fold symmetry is consistent with the Cs$^+$ adsorbed in a hollow site, such that for three rotamers the water molecules are imaged higher in their on-top positions than in the hollow sites of the other three rotamers.

The considerably larger size of the second most common species suggests the presence of a second solvation shell. The second shell exists for the triply- but not the doubly-solvated Cs$^+$ according to the optimized adsorption structure \cite{penschke_2023}, SI. It is coordinated via two hydrogen bonds to the two inner water molecules (Fig.~\ref{fig:fig 6}d, blue shaded rhombus). The six rotamers of the triply-solvated Cs$^+$ largely resemble the imaged structure. Thereby, the two higher-imaged protrusions in the interior of the flower-like structure are situated above the water molecules with a hydrogen atom pointing away from the surface. Such water molecules are known to be imaged at a larger apparent height than those with the hydrogen bonds in parallel to the surface \cite{morgenstern_2002,carrasco_2009}.
The larger apparent height along the main scanning directions is related to the tip-solvatomer interaction that is necessary to image a sub-structure within the rotamers.

The STM images suggest a system composed of singly- and triply-solvated Cs$^+$ without doubly-solvated Cs$^+$. It is supported by calculations published earlier \cite{penschke_2023}, SI. A singly- together with a triply-solvated Cs$^+$ are by 30~meV lower in energy than two doubly-solvated Cs$^+$. Hence, the sharp transition between the lifetimes (Fig.\ \ref{fig:fig 3}c) is related to the absence of doubly-solvated Cs$^+$.
Furthermore, the transition from singly-solvated to triply--solvated is accompanied by the beginning of a formation of a water network because of the hydrogen bonding within the solvation shell of the triply-solvated Cs$^+$-solvatomers.
From the electron transfer dynamics of the hydrated Cs$^+$ ions on Cu(111), we can conclude that the ion-solvent interaction at the interface is microscopically determined by a competition between Cs$^+$-water and water-water interactions. For more than two water molecules per ion, the hydrogen bond between the water is dominant and reduces the energy transfer to the solvent by 50\% compared to a single water molecule. Since we observe a weaker energy transfer and longer lifetime for larger $\rho$ we conclude that the local water network in these hydrated Cs$^+$ clusters on Cu(111) hinders dipolar rearrangement due to an increased water-water coordination compared to small $\rho$ where less water molecules without formation of a network are present.  In the present study, the combination of nanoscale structural analysis and a correlation of the energy stabilization with the vibrational frequencies of the water molecules on the Cs$^+$/Cu(111) interface facilitates the identification of H-bonding as a dominant factor in electron transfer dynamics.

\section {Conclusion}

We conclude that the number of water molecules solvating an ion on a metal surface is decisive in understanding the electron transfer dynamics across such interfaces. We find that a single water molecule attached to Cs$^+$ is more mobile and exhibits a stronger response in terms of charge screening and energy gain upon electron transfer. Such a configuration exhibits a 30\% faster response compared to a hydrogen bonded arrangement of three water molecules surrounding the ion. For the same reason, the energy gain upon solvation is larger for the single water molecule. We furthermore found that two water per ion do not occur on the investigated surface. Potentially, a suitable choice of solvent mixtures in which a low-concentration component is attached to the ion to generate energy transfer to the solvent, while the major fraction acts as a screening dielectric, will facilitate control of electron transfer rates across interfaces.

\begin{acknowledgments}
Funding by the Deutsche Forschungsgemeinschaft (DFG, German Research Foundation) under Germany's Excellence Strategy - EXC 2033 - 390677874 - RESOLV, EXC 2008/1-390540038 - UniSysCat, and by Project-ID 278162697 - SFB 1242. Furthermore, we acknowledge funding by European Union through the Horizon 2020 research and innovation program under the Marie Sklodowska-Curie Grant Agreement No. 801459 - FP - RESOMUS. We would like to thank P. Saalfrank for fruitful discussion.
\end{acknowledgments}

\end{document}